# Flexocoupling-induced soft acoustic mode and the spatially-modulated phases in ferroelectrics

Anna N. Morozovska[1,2*], Maya D. Glinchuk[3], Eugene A. Eliseev[4] and Yulian M. Vysochanskii[4†]

[1] *Institute of Physics, National Academy of Sciences of Ukraine,*
*46, Prospekt Nauky, 03028 Kyiv, Ukraine*

[2] *Bogolyubov Institute for Theoretical Physics, National Academy of Sciences of Ukraine,*
*14-b Metrolohichna str. 03680Kyiv, Ukraine*

[3] *Institute for Problems of Materials Science, National Academy of Sciences of Ukraine,*
*3, Krjijanovskogo, 03142 Kyiv, Ukraine*

[4] *Institute of Solid State Physics and Chemistry, Uzhgorod University,*
*88000 Uzhgorod, Ukraine*

**Abstract**

Using the Landau-Ginzburg-Devonshire theory and scalar approximation, we derived analytical expressions for the singular points (zeros, complex ranges) of the acoustic phonon mode (A-mode) frequency $\omega(k)$ in dependence on the wave vector $k$ and examined the conditions of the soft A-modes appearance in a ferroelectric depending on the magnitude of the flexoelectric coefficient $f$ and temperature $T$.

We predict that if the magnitude of the flexo-coefficient $f$ is equal to the temperature-dependent critical value $f^{cr}(T)$ at the temperature $T = T_{IC}$, the A-mode frequency tends to zero at $k = k_0^{cr}$ and the spontaneous polarization becomes spatially modulated in a temperature range $T < T_{IC}$. If the magnitude of the flexo-coefficient is more than the critical value in the incommensurate spatially modulated phase (SMP), the A-mode becomes zero for two wave vectors $k = k_{1,2}^{cr}$, and does not exist in the range of wave vectors $k_1^{cr} < k < k_2^{cr}$.

The comparison of calculated physical properties with measured ones are performed for some ferroelectrics with SMP phases. In particular, temperature dependence of the calculated direct and inverse static dielectric susceptibility is in an agreement with experimental data in $Sn_2P_2(Se_xS_{1-x})_6$ that gives us additional background to predict flexo-coupling induced soft acoustic "amplitudon-type" mode in the SMP phase. The available experimental data on neutron scattering in organic incommensurate ferroelectric $(CH_3)_3NCH_2COO \cdot CaCl_2 \cdot 2H_2O$ are in a semi-quantitative agreement with our theoretical results. To quantify the theory, it is necessary to measure the frequency dependence of the A-mode in a uniaxial ferroelectric with a spatially modulated phase in the temperature interval near its occurrence.

---

[*] anna.n.morozovska@gmail.com

[†] vysochanskii@gmail.com

# I. INTRODUCTION

Investigation of dynamic characteristics of phase transitions in ferroics attracts the attention of scientists for many years as the source of valuable information for fundamental physics and technical applications [1, 2] supposed that any phase transition leads to instability of some phonon vibration. In particular for ferroelectrics the frequency $\omega_{TO}$ of transverse optic (TO) soft mode depends on temperature $T$ and $\omega_{TO}(T_C) = 0$ at transition temperature $T = T_C$. Therefore static displacements of atoms at the phase transition correspond to frozen displacements of soft phonon vibration mode.

Inelastic neutron scattering experiments proved that not only the TO mode softens substantially with decreasing temperature to freeze out at $T_C$ in ferroics (such as ferroelectric perovskites), and finite wave vector anomalies appear in the related transverse acoustic (TA) mode [3, 4, 5, 6]. Acoustic soft mode that rules the antiferrodistortive phase transition was observed in incipient ferroelectric $SrTiO_3$ [7]. The pronounced softening of TA mode, as remarkable lowering of hypersound velocity for the transverse-acoustic phonons, was observed in the incommensurate ferroelectric $Sn_2P_2(Se_xS_{1-x})_6$ in the vicinity of Lifshitz point [8]. Besides original papers, lots of information about experimental and theoretical investigations of soft modes in different ferroics can be found in the seminal textbooks [1, 9, 10], however the question about existence of soft acoustic modes in the incommensurate spatially modulated phases (**SPM**) of ferroics was very little investigated.

Basic experimental methods, which contain information about the soft modes and spatial modulation of the order parameter in ferroics (such as antiferroelectrics, proper and incipient ferroelectrics) are dielectric measurements [11], inelastic neutron scattering [5, 12, 13, 14, 15, 16, 17], X-ray [18, 19, 20, 21], Raman [22] and Brillouin [8, 18, 21, 23, 24, 25] scatterings and ultrasonic pulse-echo method [8-24] allowing hypersound spectroscopic measurements.

Calculations of the spatial modulation in the form of polarization and strain gradients used to be performed in the first approximation of the self-consistent field theory. This approach allowed considering correlation effects [10] as well as flexoelectric effect existing in a medium of arbitrary symmetry [26, 27, 28]. Direct and converse static flexoelectric effects, which leads to the appearance of polarization due to the strain gradient and vise versa [27, 28], influences strongly many physical properties of macrosized and nanoscale ferroics, such as domain walls structure, surface and interface states, correlation radius and many others [29, 30, 31].

It is obvious that wave excitation of any nature is impossible without a local gradient of the corresponding physical quantity [32, 33]. For example, an acoustic wave will be inevitably accompanied by a local gradient of mechanical strains and stresses. Due to the direct static flexoelectric effect, the wave of the strain gradient induces the wave of the electric polarization even in paraelectric phase (i.e. local polarization, the mean value of which is zero). The local polarization gradient, due to the converse flexoelectric effect, will affect the elastic stresses associated with the wave. Moreover, the static flexocoupling can induce incommensurate SMP [8, 34, 35, 36] in ferroics. The dynamic flexoeffect originated from polarization and strain time derivatives. The notion about dynamic flexoelectric coupling [37, 38, 39] impact on their phonon spectra has been absent until recently [40, 41]. Thus, the static and dynamic flexoelectric effects should influence the propagation of acoustic waves in all solids, although the effect should be more pronounced for short than for long wavelengths [42].

Recently using Landau-Ginzburg-Devonshire (**LGD**) approach Morozovska et al. [40, 41] considered the influence of the flexocoupling on the appearance of SMP, and on the properties of optic and acoustic phonons in the ferroelectric and paraelectric phases of ferroelectrics $PbTiO_3$, $Sn_2P_2(S,Se)_6$, and paraelectric $SrTiO_3$. In order to derive the analytical expressions from the LGD free energy, which includes the higher gradient terms, static and dynamic flexocoupling, we were subjected to restrict our consideration by one component of the polarization and strain. These restrictions can be called the **scalar approximation** [41]. The consequence of this approximation is the occurrence of only one optical and one acoustic mode, which interact via electrostriction and flexoelectric couplings.

Allowing for we are interested in SMP phase transition that does not belong to ferroelectric one [43], so that optic modes are not suitable for SMP transition in what follows we are going to consider mainly acoustic mode dispersion and temperature dependence.

The simplified dispersion law cannot describe the interaction between different transverse and longitudinal optical modes and three acoustic modes induced by cooperative effects, flexoelectric and electrostriction couplings in the ferroelectric phase of multiaxial ferroelectrics. Moreover, the coupling between different optical phonon modes can either act cooperatively with the flexocoupling or against it (see e.g., the models by Kappler and Walker [44], and Hlinka et al [45] adopted for organic ferroelectrics such as $(CH_3)_3NCH_2COO \cdot CaCl_2 \cdot 2H_2O$). If acoustic and optical modes are considered, their mutual coupling contributes to the SMP appearance similarly to the flexoelectricity standing alone [45]. Actually, it was shown that elastic softening in the hypersound range in $Sn_2P_2(Se_xS_{1-x})_6$ is induced mostly by linear interaction between soft optic and acoustic phonon branches and Landau-Khalatnikov model explains temperature dependence of hypersound velocity in the ferroelectric phase [8].

Taking into account the limitations of the scalar approximation validity, we can reasonably apply the analytical results to describe the experimentally observed phonon dispersion in uniaxial ferroics only, e.g., for the monoclinic ferroelectrics $Sn_2P_2(Se_xS_{1-x})_6$ [46, 47], as well in the paraelectric phase of organic ferroelectrics like $(CH_3)_3NCH_2COO \cdot CaCl_2 \cdot 2H_2O$ [45] if the interaction between different phonon modes appeared relatively small at high temperatures.

**Motivated by above argumentation**, in this work we derived analytical expressions for the singular points (zero points, complex ranges) of the acoustic phonon mode frequency $\omega(k)$ in dependence on the wave vector $k$ and examined the conditions of the soft acoustic modes appearance in ferroelectrics depending on the magnitude of the flexoelectric coefficient $f$ and temperature $T$ of the ferroelectric. It was shown that acoustic mode becomes soft only at $|f| > f_{cr}$, when incommensurate spatially modulated phase appears. Temperature dependence of dielectric susceptibility in this phase was calculated and the slight fracture at phase transition temperature is in agreement with experiment, at that the fracture instead of maximum at the phase transition temperature being the characteristic feature of any SMP transitions [43]. Comparison of calculated and measured dispersion of acoustic mode frequency $\omega(k)$ in abovementioned organic ferroelectric shows that our theory quantitatively describes the experiment [45].

## II. ANALYTICAL SCALAR THEORY

Using the Landau-Ginzburg-Devonshire (LGD) theory and scalar approximation in the simplest one-component 1D case considered hereinafter, the bulk part of the free energy $F$ of ferroelectric, which depends on polarization component $P$, and strain component $u$, and their gradients, has the following form [41]:

$$F_V[P,u] = \int dx \left( \begin{array}{c} \dfrac{\alpha(T)}{2}P^2 + \dfrac{\beta}{4}P^4 + \dfrac{\gamma}{4}P^6 + \dfrac{g}{2}\left(\dfrac{\partial P}{\partial x}\right)^2 - PE \\ - quP^2 + \dfrac{c}{2}u^2 + \dfrac{v}{2}\left(\dfrac{\partial u}{\partial x}\right)^2 - \dfrac{f}{2}\left(P\dfrac{\partial u}{\partial x} - u\dfrac{\partial P}{\partial x}\right) \end{array} \right) \quad (1)$$

According to Landau theory [48, 49], the coefficient $\alpha$ linearly depends on the temperature $T$ for proper ferroelectrics, $\alpha(T) = \alpha_T(T - T_C)$. $T_C$ is the Curie temperature. All other coefficients in Eq.(1) are supposed to be temperature independent. Coefficient $\beta>0$ for the ferroelectrics with the second order phase transition, and $\beta<0$ for the first order one. Nonlinear stiffness $\gamma$ should be non negative ($\gamma \geq 0$) for the stability of the functional (1) for all $P$ values. The gradient coefficients $g>0$ and $v>0$ determine the magnitude of the gradient energy. The polarization interacts with an external electric field $E$. Since we did not include the term $PE_d[P]/2$ in Eq.(1) we assume that the depolarization field is absent, and so can consider only transverse fluctuation of polarization for which $div[P] = 0$. The case corresponds to the transverse variation of polarization components. The electrostriction coefficient $q$ can be positive or negative. The elastic stiffness $c$ and the strain gradient coefficient $v$ should be always positive for the functional stability. Coefficient $f$ is the component of the static flexocoupling tensor, whose sign is not fixed.

Unlike our previous work [41], we did not included the higher polarization gradient term proportional to $w(\partial^2 P/\partial x^2)^2$ in Eq.(1) and consider only positive $g>0$. This is done in order to study those and only those types SMP, which originated with the increase of flexocoupling strength $f$ only, due the presence of Lifshitz invariant $\frac{f}{2}\left(P\frac{\partial u}{\partial x} - u\frac{\partial P}{\partial x}\right)$ in Eq.(1). The higher polarization gradient term determines the possible appearance of the minima on the optic mode frequency and so conditions the appearance of incommensurate phase with increase of its strength $w$ (see e.g. figs.1-2 in review [50]). We restrict our consideration by $w=0$, because we regard that it is really important to distinguish between the scenario of SPM origin in initially homogeneous commensurate ferroics (with $g>0$ and $w=0$) under the flexocoupling strength $f$ increase [35] from other possible scenarios of incommensurate SPM appearance in different ferroics (see Ref. [50] and refs. therein).

The Lagrange function $L = \int_t dt(F - K)$ consists of the free energy $F$ given by Eq.(1) and the kinetic energy

$$K = \int dx \left( \frac{\mu}{2}\left(\frac{\partial P}{\partial t}\right)^2 + M \frac{\partial P}{\partial t} \frac{\partial U}{\partial t} + \frac{\rho}{2}\left(\frac{\partial U}{\partial t}\right)^2 \right), \qquad (2)$$

which includes the dynamic flexocoupling [37, 38] with the magnitude $M$; $\rho$ is the density of a material; $\mu$ is a kinetic coefficient. The elastic displacement component $U$ is related with the strain $u$ as $u = \partial U/\partial x$.

The dependence of the soft phonon frequency $\omega$ on its wave vector $k$, i.e. the dispersion law $\omega(k)$, can be calculated from the time-dependent dynamic equations of state for the polarization and elastic displacement components $P$ and $U$, correspondingly [51]. The dynamic equations of state are obtained from the variation of the Lagrange function $L = \int_t dt(F - K)$ on $P$ and $U$, where the free energy $F$ is given by Eq.(1) and the kinetic energy $K$ is given by Eq.(2). Let us find the solution of these equations after their linearization in the vicinity of spontaneous values $P_S$ and $u_S$. The presentation of linearized solution in the form of the Fourier integral is:

$$P = P_S + \int dtdk\, \exp(ikx - i\omega t)\widetilde{P}(\omega, k), \quad u = u_S + \int dtdk\, \exp(ikx - i\omega t)\widetilde{u}(\omega, k). \qquad (3)$$

Homogeneous spontaneous strain and order parameter are $P_S^2 = \frac{1}{2\gamma}\left(\sqrt{\beta^{*2} - 4\alpha\gamma} - \beta^*\right)$ and $u_S = \frac{q}{c}P_S^2$, where the coefficient $\beta^* = \left(\beta - 2\frac{q^2}{c}\right)$. The perturbation electric field $E$ has the Fourier expansion $E = \int dk \exp(ikx - i\omega t)\widetilde{E}(\omega, k)$. Using the methodology described in details in Ref.[41], we obtained the linearized solution of Euler-Lagrange equations for the perturbations in the form]:

$$\widetilde{U} = -\frac{fk^2 + 2ikqP_S - M\omega^2}{vk^4 + ck^2 - \rho\omega^2}\widetilde{P}, \qquad \widetilde{P} = \widetilde{\chi}(k,\omega)\widetilde{E} \qquad (4)$$

The linear dynamic susceptibility $\widetilde{\chi}(k,\omega)$ introduced in Eq.(4) is given by the expression:

$$\widetilde{\chi}(k,\omega) = \frac{vk^4 + ck^2 - \rho\omega^2}{(\alpha_S + gk^2 - \mu\omega^2)(vk^4 + ck^2 - \rho\omega^2) - (fk^2 - M\omega^2)^2 - 4k^2q^2P_S^2}. \qquad (5)$$

Hereinafter the positive temperature-dependent function $\alpha_S(T)$ is introduced:

$$\alpha_S(T) = \alpha(T) + \left(3\beta - 2\frac{q^2}{c}\right)P_S^2(T) + 5\gamma P_S^4(T). \qquad (6)$$

## III. SOFT ACOUSTIC MODE IN THE SPATIALLY MODULATED PHASE
### A. Analytical expressions for the soft phonon spectra

Using the methodology described in details in Ref.[40, 41], from the singularity of the generalized susceptibility (corresponding to zero points of Eq.(5) denominator) we derived the characteristic equation for the frequency $\omega(k)$, which form as a power expansion on $\omega(k)$ is the following:

$$(\mu\rho - M^2)\omega^4 - C(k)\omega^2 + B(k) = 0, \qquad (7a)$$

wherein the functions $C(k)$ and $B(k)$ are introduced

$$C(k) = \alpha_S\rho + (c\mu - 2fM + g\rho)k^2 + \mu v k^4, \qquad (7b)$$

$$B(k) = k^2\left(\alpha_S c - 4q^2 P_S^2 + (cg + \alpha_S v - f^2)k^2 + gvk^4\right). \qquad (7c)$$

The solution of biquadratic Eq.(7a) can be represented in the form:

$$\omega_{1,2}^2(k) = \frac{C(k) \pm \sqrt{C^2(k) - 4(\mu\rho - M^2)B(k)}}{2(\mu\rho - M^2)}, \qquad (8)$$

Dispersion relation (8) contains one optical (**O**) and one acoustic (**A**) phonon modes, which corresponds to the signs "+" and "−" before the radical, respectively. The **O**-mode is in fact transverse, and the **A**-mode can be longitudinal or transverse. The "gap" between these modes is proportional to the value $\dfrac{\sqrt{C^2(k) - 4(\mu\rho - M^2)B(k)}}{2(\mu\rho - M^2)}$.

The series expansion of the **O**-mode frequency $\omega(k)$ for the small $k \to 0$ gives $\omega \approx \sqrt{\frac{\alpha_S \rho}{\mu\rho - M^2}}(1 + O(k^2))$. Hence the dependence of the O-mode on the flexocouling constant is weak and non-critical at small $k$. The softening law of the optical phonons is valid near Curie temperature $T_C$, $\omega(T) \sim \sqrt{\alpha_S(T)} \sim \sqrt{|T_C - T|}$.

Note that the flexocoupling impact leads to the fact that the condition $\omega^2 = 0$ can be valid for A-mode not only at $k=0$. Actually the condition $\omega^2 = 0$ leads to the simplification in Eq.(4a), namely it reduced to the biquadratic equation for $k = k_{cr}$,

$$k_{cr}^2 \left((\alpha_S + gk_{cr}^2)(vk_{cr}^2 + c) - f^2 k_{cr}^2 - 4q^2 P_S^2\right) = 0. \tag{9}$$

Except for the trivial solution $k_{cr} = 0$, the biquadratic equation (6) has four roots, $k_{cr} = \pm k_{1,2}^{cr}$:

$$k_{1,2}^{cr} = \sqrt{\frac{c}{2v}\left(\frac{f^2}{cg} - 1 - \frac{\alpha_S v}{cg} \mp \sqrt{\left(\frac{f^2}{cg} - 1 - \frac{\alpha_S v}{cg}\right)^2 - 4\frac{v}{cg}\left(\alpha_S - 4\frac{q^2}{c}P_S^2\right)}\right)}. \tag{10}$$

Subscripts "1" and "2" corresponds to the signs "−" and "+" before the radical, respectively.

Since the condition $(\alpha_S - 4(q^2/c)P_S^2) \geq 0$ follows from the definition of $\alpha_S > 0$ and functional stability condition, both roots (7) are real and so exist under the condition $|f| > f_{cr}$, where the temperature-dependent critical value $f_{cr}$ is given by expression:

$$|f_{cr}(T)| = \sqrt{\alpha_S(T)v + cg + 2\sqrt{cgv\left(\alpha_S - 4\frac{q^2}{c}P_S^2(T)\right)}}. \tag{11}$$

Under the condition $|f| = f_{cr}$ both roots for A-mode in Eq.(10) coincides, $k_1^{cr} = k_2^{cr} = k_0^{cr}$ and $\omega(k_0^{cr}) = 0$ at $\left(\frac{f^2}{cg} - 1 - \frac{\alpha_S v}{cg}\right) = 2\sqrt{\frac{v}{cg}\left(\alpha_S - 4\frac{q^2}{c}P_S^2\right)}$, where the critical value is given by expression:

$$k_0^{cr} = \sqrt{\frac{c}{2v}\left(\frac{f^2}{cg} - 1 - \frac{\alpha_S v}{cg}\right)} \tag{12}$$

**B. Flexocoupling-induced SMPs**

The critical point $\omega_2(k_0^{cr}) = 0$ corresponds to the system transition at temperature $T = T_{IC}$ from the paraelectric phase (**PE**) into the incommensurate spatially-modulated phase (**SMP**) with polarization $P(x) = \sum_{i=1}^{2} \delta P_i \sin(k_i^{cr} x)$ that gradually tends to the homogeneous ferroelectric (**FE**) phase with temperature decrease to $T = T_C$ [see Ref.[41] and the regions of PE and SMP phase in **Fig.1(a)**].

Under the condition $P_S^2 = 0$, that is valid in the PE phase at the onset of SMP phase, the coefficient $\alpha_S = \alpha_T(T_{IC} - T_C)$ and so the critical value is $k_0^{cr} = \sqrt{\dfrac{f^2 - cg - v\alpha_T(T_{IC} - T_C)}{2vg}} \equiv \sqrt[4]{\dfrac{c\alpha_T(T_{IC} - T_C)}{vg}}$, i.e. $k_0^{cr} \sim (T_{IC} - T_C)^{1/4}$. Three phases PE, FE and SMP coexist at the flexocoefficient value $|f| = \sqrt{cg}$ that is a formal analog of a critical Liftshitz point (**LP**).

The transition temperature from PE to the SMP phase, $T_{IC}$, can be found from the condition $B(k_0^{cr}, T_{IC}) = 0$, i.e. from the equation $(c\alpha_S(T_{IC}) - 4q^2 P_S^2 + (cg + \alpha_S v - f^2)(k_0^{cr})^2 + gv(k_0^{cr})^4) = 0$. Under the condition $P_S^2 = 0$ (valid in the paraelectric phase at the onset of SMP phase) the coefficient $\alpha_S = \alpha_T(T - T_C)$ and so the latter equation for $T_{IC}$ reduces to the condition of zero inner determinant in Eq.(10) at $T = T_{IC}$, namely $f^2 = \alpha_S(T_{IC})v + cg + 2\sqrt{cgv\alpha_S(T_{IC})} \equiv \left(\sqrt{\alpha(T_{IC})v} + \sqrt{cg}\right)^2$. The solution of the equation exists at flexocoefficients $|f| > \sqrt{cg}$ and has the form:

$$T_{IC}(f) = T_C + \dfrac{(|f| - \sqrt{cg})^2}{\alpha_T v}. \tag{13}$$

Expression (13) shows how the transition temperature $T_{IC}(f)$ to the SMP phase depends on the flexocouling constant $f$ and the elastic strain gradient constant $v$ [see **Fig.1(b)**]. Actually the difference $(T_{IC}(f) - T_C)$ is directly proportional to the second power of the flexocoefficient $(|f| - \sqrt{cg})^2$ and inversely proportional to the strain gradient coefficient $v$, i.e. $(T_{IC}(f) - T_C) \sim (|f| - \sqrt{cg})^2 / v$.

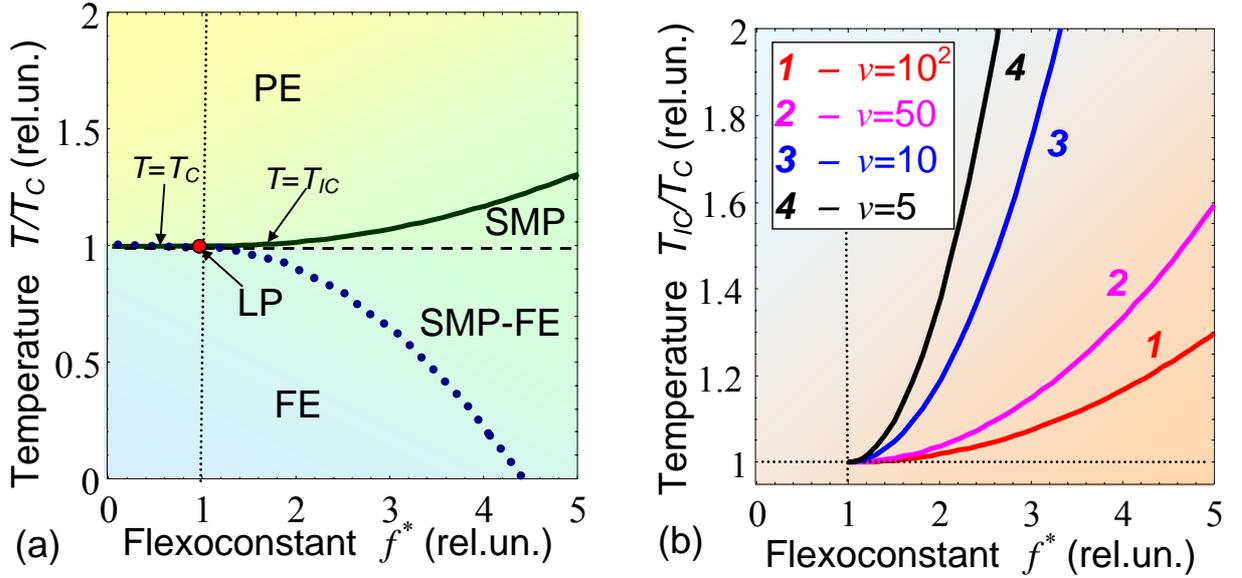

**FIG. 1. (a)** Phase diagram in coordinates relative temperature $T/T_C$ - absolute value of dimensionless flexoconstant $f^* = |f|/\sqrt{cg}$ calculated for $v = 10^{-7}$V s$^2$/m$^2$ and other parameters corresponding to the solid solution Sn$_2$P$_2$(S,Se)$_6$ [listed in **Table I**]. The boundaries between the paraelectric (**PE**), homogeneous ferroelectric (**FE**), incommensurate spatially modulated phase (**SMP**) that gradually tends to the homogeneous FE phase via partially polar modulated phase (**SMP-FE**) are shown by solid, dashed and dotted curves, respectively. Red circle indicates the Liftshitz point (**LP**), where the PE, FE and SMP phases coexist. **(b)** Dependence of the SMP phase transition temperature $T_{IC}/T_C$ on the dimensionless flexoconstant $f^*$ calculated for different $v$ values, $v = 100$ (red curves 1), $v = 50$ (magenta curve 2), $v = 10$ (blue curve 3) and $v = 5$ in $10^{-9}$V s$^2$/m$^2$ (black curve 4).

Also we predict that the polar incommensurate modulation of polarization, $P(x) = P_0(T) + \sum_{i=1}^{2} \delta P_i \sin(k_i^{cr} x)$, can appear at temperatures below $T_C$ and high enough flexocoefficients $|f| > \sqrt{cg}$ in the initially commensurate ferroelectric [see the region of **SMP-FE** phase in **Fig.1(a)**]. The only physical reason of the SMP-FE phase existence in the considered case is the flexocoupling, since we impose the conditions $g>0$ and $w=0$ in the free energy (1) (i.e. do not include the gradient term proportional to $w(\partial^2 P/\partial x^2)^2$). However we included the higher term $\gamma P^6$, keeping in mind that the strength $\gamma$ can strongly affect on the SMPs stability via the higher harmonics and modulation pattern nonlinearities [52, 53]. The analytical expression for the transition temperature from incommensurate SMP-FE phase to commensurate homogenous FE is absent, at that the diffuse boundary between the phases can be found numerically from the minimum of the free energy (1).

The superposition of SMP and FE phases was named "rippled phases", and it was regarded to appear in the incommensurate ferroics of **type I** (see e.g. Refs. [54, 50] for classification). In fact, the existence of the SMP-FE phase in the incommensurate ferroics of **type I** (corresponding to the second order phase transition from almost solitonic incommensurate SPM to FE phase in incipient ferroelectrics like $K_2SeO_4$) or in the **type II** (corresponding to the first order phase transition from almost sinusoidal incommensurate SPM to FE phase in proper ferroelectrics like $Sn_2P_2(Se_xS_{1-x})_6$ family) can be crucial for our model verification, since our prediction of SPM-FE phase is based on the flexocoupling impact only. Allowing for the presence of the almost symmetric maxima of dielectric permittivity at $T_{IC}$, the experimental data speaks in favor of the rippled phase existence in $Sn_2P_2(Se_xS_{1-x})_6$ near the boundary of normal FE phase (see **Fig. 1** in Ref.[55]).

**C. Soft acoustic mode behavior in the vicinity of critical wave vectors**

As a next step let us expand the frequency $\omega(k)$ of **A**-mode in series for the small $k \to 0$ and in the vicinity of critical values, $k \to \pm k_{1,2}^{cr}$. Corresponding solution of Eq.(8) at small $\omega$ has the form:

$$\omega^2 = k^2 \frac{(\alpha_S c - 4q^2 P_S^2 + k^2(\alpha_S v + cg - f^2) + gvk^4)}{(\alpha_S \rho + k^2(c\mu + g\rho - 2fM) + \mu v k^4)} \equiv \frac{gvk^2(k^2 - (k_1^{cr})^2)(k^2 - (k_2^{cr})^2)}{\alpha_S \rho + k^2(c\mu + g\rho - 2fM) + \mu v k^4} \quad (14)$$

From Eq.(14) we obtained that $\omega^2 \approx k^2 \left(\frac{c}{\rho} - \frac{4q^2 P_S^2}{\alpha_S \rho}\right)$ at small $k$; and

$$\omega^2 \approx \frac{2gv(k_1^{cr})^3((k_2^{cr})^2 - (k_1^{cr})^2)(k_1^{cr} - k)}{\alpha_S \rho + (k_1^{cr})^2 (c\mu + g\rho - 2fM) + \mu v (k_1^{cr})^4} \quad \text{at } k \to k_1^{cr} - 0. \text{ Similar speculations are valid in the case}$$

$k \to k_2^{cr} + 0$. That say, the expressions for the frequency is valid:

$$\omega(k \to 0) \approx k \sqrt{\frac{c}{\rho} - \frac{4q^2 P_S^2}{\alpha_S \rho}} \left(1 + \lambda \frac{k^2}{2}\right), \quad (15a)$$

$$\omega(k \to k_1^{cr} - 0) \approx \frac{\sqrt{2gv(k_1^{cr})^3 |(k_1^{cr})^2 - (k_2^{cr})^2|} \sqrt{k_1^{cr} - k}}{\sqrt{\alpha_S \rho + (k_1^{cr})^2 (c\mu + g\rho - 2fM) + \mu v (k_1^{cr})^4}}, \quad (15b)$$

$$\omega(k \to k_2^{cr} + 0) \approx \frac{\sqrt{2gv(k_2^{cr})^3 |(k_1^{cr})^2 - (k_2^{cr})^2|} \sqrt{k - k_2^{cr}}}{\sqrt{\alpha_S \rho + (k_2^{cr})^2 (c\mu + g\rho - 2fM) + \mu v (k_2^{cr})^4}}, \quad (15c)$$

Where the small constant $\lambda$ is temperature-dependent [56]. Note that $\left|(k_1^{cr})^2 - (k_2^{cr})^2\right| \equiv \left|\frac{\alpha_S v + cg - f^2}{gv}\right|$ in accordance with Vieta theorem.

In the "paraelectric" vicinity of the transition to SMP phase, i.e. at $T \to T_{IC} + 0$, and wave vectors $k \to k_0^{cr}$, the frequency of the A-mode given by Eq.(14) can be estimated as following:

$$\omega\left(k \to k_0^{cr}\right) \approx \frac{2\sqrt{gv}\left(k_0^{cr}\right)^2 \left|k - k_0^{cr}\right|}{\sqrt{\alpha_s(T)\rho + (c\mu + g\rho - 2fM)\left(k_0^{cr}\right)^2 + \mu v\left(k_0^{cr}\right)^4}} \quad (16)$$

The approximate temperature dependence given by Eq.(16) can be simplified by using Eq.(12) at $P_S^2 = 0$, when $k_0^{cr} \equiv \sqrt[4]{c\alpha_T(T_{IC} - T_C)/vg}$ and $T \to T_{IC}$, the acoustic frequency $\omega\left(k \to k_0^{cr}\right) \sim \sqrt{\frac{c}{\rho}}\left|k - k_0^{cr}\right|$.

**Figure 2** illustrates the graphical sense of the expressions (15)-(16). The frequency of the **A**-mode is independent on the flexocoupling at very small $k$, namely $\omega(k) \cong \left(\sqrt{c/\rho}\right)k$ in accordance with Eq.(15a). If the inequality is valid $|f| \ll f_{cr}$ **A**-mode frequency is equal to zero at $k=0$ only, and it monotonically increases with $k$ increase (curve 1). With $f$ increase in the range $f_{inf}(T) < |f| < f_{cr}(T)$ **A**-mode frequency is zero at $k=0$ only, but then the local minimum appears at $k = k_{min}$ (curve 2). The values $f_{inf}(T)$ and $k_{min}$ can be determined from Eqs.(8) for the conditions $\frac{d\omega_2(k)}{dk} = 0$ and $\frac{d^2\omega_2(k)}{dk^2} \geq 0$. Analytical calculation of $f_{inf}(T)$ and $k_{min}$ is very cumbersome however numerically it is easy to do.

Under the condition $|f| = f_{cr}$ the **A**-mode frequency is also zero at $k = k_0^{cr}$ (namely its graph touches the k-axis, see curve 3), where $k_0^{cr}$ is given by Eq.(12). Under the condition $|f| > f_{cr}$ the **A**-mode frequency is positive in the regions $0 < k < k_1^{cr}$ and $k > k_2^{cr}$, it is zero at $k=0$, $k = k_1^{cr}$ and $k = k_2^{cr}$, and does not exist in the region $k_1^{cr} < k < k_2^{cr}$ (curve 4). At that the squire root laws $\omega(k) \sim \sqrt{k_1^{cr} - k}$ and $\omega(k) \sim \sqrt{k - k_2^{cr}}$ are valid in the vicinity of critical values $k \to k_1^{cr} - 0$ and $k \to k_2^{cr} + 0$, respectively (see dashed curves).

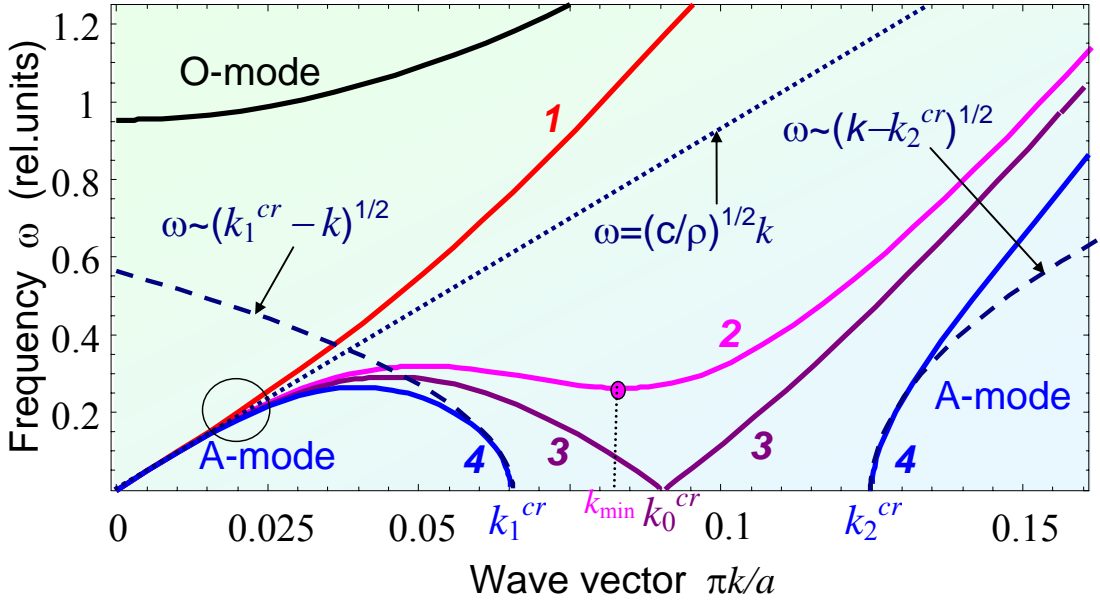

**FIG. 2.** Dependences of the phonon frequency $\omega^*$ on the wave vector $k^* = ak/\pi$. **O-mode** (black curve) is not sensitive to the f value for small $k$. **A-mode** curves calculated at $f = 0$ (red curve 1), $|f| = 0.5 f_{cr}$ (magenta curve 2), $|f| = f_{cr}$ (dark violet curve 3) and $|f| \approx 1.1 f_{cr}$ (blue curves 4) are shown. Dotted curve is the linear approximation (15a), dashed curves show are squire root approximations (15b) and (15c), respectively.

Analyzing the results shown in **Fig.2** for the case $|f| > f_{cr}$, one should ask two reasonable questions. What happens in a ferroelectric if the spatial fluctuations of the wave vector k are within the range $k_1^{cr} \leq k \leq k_2^{cr}$? Are there any experiment that reveals the "softening" of the A-mode similar to the ones shown by the curves 2, 3 and 4 in **Fig.2**?

The answer to the first question is quite simple. Using results of Refs.[40, 41], the spatially modulated phase with the modulation vectors $k = k_1^{cr}$ and $k = k_2^{cr}$ can occurs in the temperature range for which $|f| > f_{cr}(T)$. In essence, this means that any fluctuation having a wave vector in the interval $k_1^{cr} \leq k \leq k_2^{cr}$ relaxes in a short time to the value $k = k_1^{cr}$ or to $k = k_2^{cr}$ (which is less probable, since the larger is the wave vector, the greater is the influence of the lattice anharmonicity on the phonon spectrum, and so the less applicable are the linear approximation in the perturbations of $\tilde{P}$ and $\tilde{u}$ Fourier integrals (3)). In this sense, acoustic vibrations with wave vectors $k_1^{cr} < k < k_2^{cr}$ are absent in the lattice, so the frequency of the corresponding A-mode is absent in this interval of wave vectors. The "normal" A-mode exists in the interval $0 < k < k_1^{cr}$ and $k > k_2^{cr}$, it softens approaching the point $k = k_1^{cr}$, as well as at $k = k_1^{cr}$.

The answer to the second question is not simple, because the value of the flexo-coefficient *f* is fixed for each specific ferroelectric, and in general case it is relatively weakly dependent on the temperature, and so one can not vary it within the necessary limits in order to "get into" the interval $|f| > f_{cr}(T)$. However, experimentalists have all the possibilities to measure phonon spectra using neutron scattering in a wide range of temperatures in order to vary $f_{cr}(T)$ that essentially depends on temperature, primarily due to the temperature dependence of $\alpha_S(T)$ according to Eq. (6). In particular, the parameter depends linearly on the temperature in the PE phase, $\alpha_S(T) = \alpha(T) \cong \alpha_T(T - T_C)$. Thus, the form of the acoustic branch can be changed similarly to the changes shown in **Fig. 2**, by changing only the temperature, i.e. the parameter $\alpha_S(T)$, for a fixed value of the flexo-coefficient $f$. A typical scenario is shown in **Fig.3(a)**. The dependence $\omega(k)$ shows only a very small bend for the highest relative temperature $\alpha_S^* = 1.4$ (curve 4), where the dimensionless parameter $\alpha_S^* = v\alpha_S/(cg)$ is introduced. With the temperature decrease corresponding to the value of $\alpha_S^* = 1.0$, a local minimum appears instead of the bend, which becomes deeper and noticeable with $\alpha_S$ decrease (curve 3). With a further decrease in temperature to a critical value of $\alpha_S^* = 0.9$ (that corresponds to the temperature of the SMP transition, $T = T_{IC}$,), the zero point $\omega(k) = 0$ arises at the at $k = k_0^{cr}$ (curve 2). Finally, at $\alpha_S^* = 0.9$, a "dip" occurs on the acoustic branch in the wave vector interval $k_1^{cr} < k < k_2^{cr}$ (curve 1) corresponding to the appearance of SMP in a certain temperature interval $T < T_{IC}$. At that dimensionless flexoconstant $f^* = |f|/\sqrt{cg}$ that controls the values $k_{1,2}^{cr}$ determined by Eq.(10). The dimensionless variables, which were essentially used in the illustrative **Figs 1-3,** are described in the Supplement.

An analysis of the available experimental data has shown that the anomalous flattening, bends, inflections, minima, and maxima in the A-mode frequency spectra $\omega(k)$ are observed experimentally quite often at $k \sim (0.2 - 0.4)\pi/a$ [8, 13, 16, 47], and are explained by the interaction of several optical and acoustic modes in proper ferroelectrics [45], as well as by electron-density multiphonon interactions in incipient ferroelectrics [6]. For example, from the experimental data shown in **Fig.3(b)** it can be seen that the bend on the lowest acoustic A-mode is absent at 300 K in organic ferroelectric $(CH_3)_3NCH_2COO \cdot CaCl_2 \cdot 2H_2O$. A noticeable minimum appears even at 205 K, and the softening of the A-mode continues at 171 K. The true incommensurate phase occurs at 164 K and coexists with commensurate FE phase from $T_C$= 125 K up to 43 K according to the "devil staircase" scenario [16]. However the experimental scenario [16] agrees qualitatively with the theoretical one shown in **Fig. 3 (a).** To illustrate this we add solid curves for A-mode calculated from Eq.(8) for the temperatures (300 – 164) K in **Fig.3(b)**. One can see the agreement between theoretical curves and experimental points.

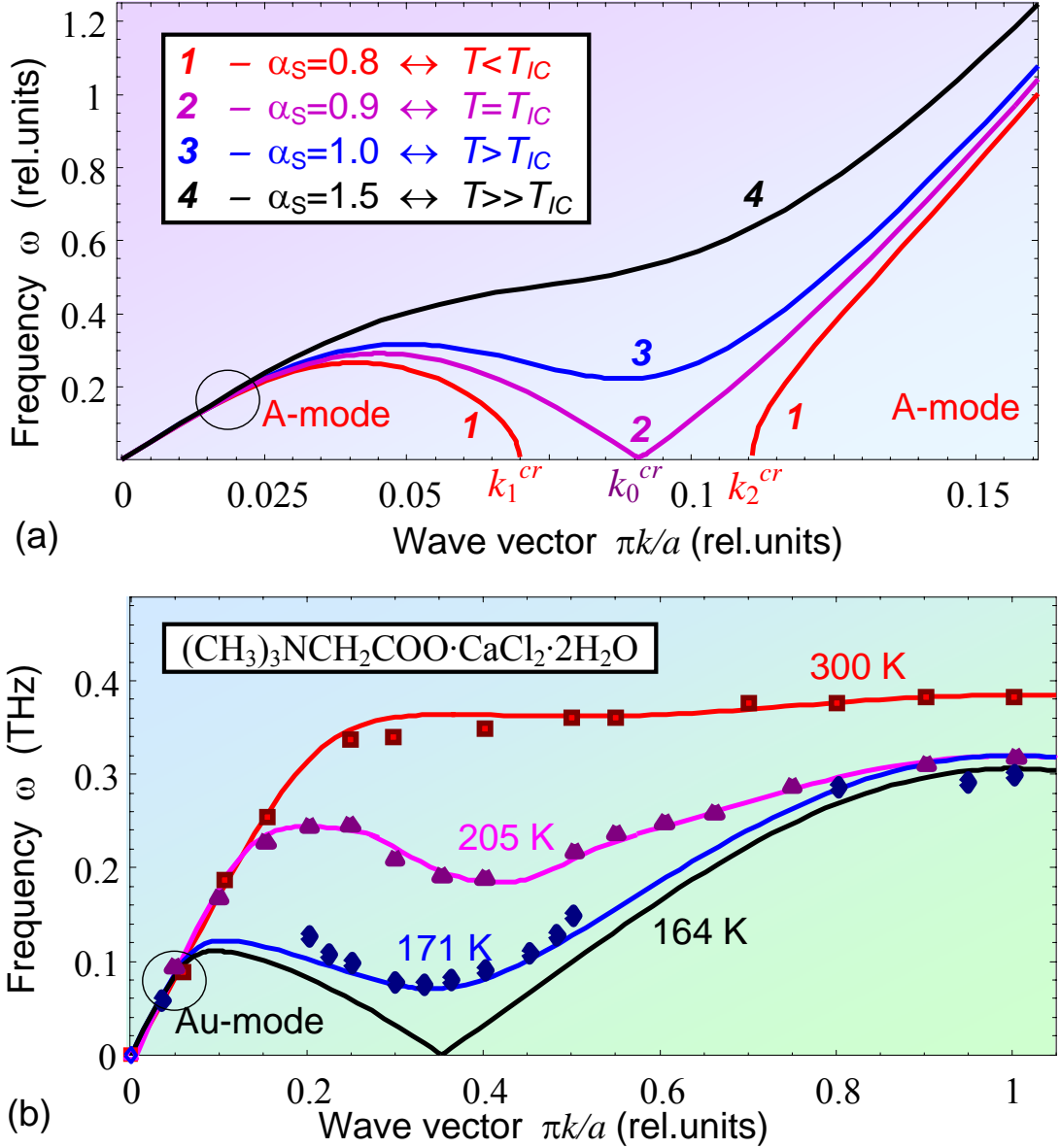

**FIG. 3.** (a) Dependences of the **A**-mode frequency $\omega^*$ on the wave vector $k^* = ak/\pi$. The curves calculated for fixed $f^* = 1.98$ relative units and different temperatures, which corresponds to different $\alpha_S^* = 0.8$ (red curves 1), $\alpha_S^* = 0.9$ (magenta curve 2), $\alpha_S^* = 1.0$ (blue curve 3) and $\alpha_S^* = 1.5$ relative units (black curve 4). **(b)** Acoustic phonon spectra measured experimentally in incommensurate organic ferroelectric $(CH_3)_3NCH_2COO \cdot CaCl_2 \cdot 2H_2O$ (symbols from Refs. [16, 45]) and calculated by us at 300, 205, 171 and 164 K (solid curves). Experimental data for 164 K is absent. Fitting parameters are $T_C$=125 K, $f^* \approx 3.14$ and $\alpha_S^*(T) = (T - T_C)/T_C$.

However the "true" soft A-modes, which are absent at $k_1^{cr} < k < k_2^{cr}$ have not been observed so far, although the absence of A-mode in proper and incipient ferroelectrics at definite k-range is possible in a number of microscopic theories (see e.g. Fig. 6(b) for organic ferroelectric (CH$_3$)$_3$NCH$_2$COO·CaCl$_2$·2H$_2$O in Ref. [45], Fig.2 for SrTiO$_3$ in Ref.[7] and Fig 8(a) for Sn$_2$P$_2$(Se$_{0.28}$S$_{0.72}$)$_6$ in Ref.[8]). It is also quite possible that the value $f_{cr}(T)$ is not reached for some real materials, or the range of critical values $k_1^{cr} < k < k_2^{cr}$ goes beyond the region of the linear model applicability. The possibility of other reasons cannot be excluded. In particular the lattice instability analysis in Sn$_2$P$_2$(S,Se)$_6$ solid solutions in the framework of a polarizable ion model shows that a possible reason for the absence of a soft acoustic mode is nonorthogonality of the spontaneous polarization vector and modulation wave vector which both lie in the monoclinic symmetry plane [25].

Let us discuss the indirect evidences of the A-mode softening in the incommensurate ferroelectrics with SPM phase. The phase velocity of acoustic wave is given by expression $V(k) = |\omega|/k$. Temperature dependences of the A-mode velocity calculated for different values of flexo-constant and temperatures $T > T_C$ ($\alpha_S^* > 0$) is shown in **Fig.4(a).** The velocity monotonically and saturates with $T$ increase at $f = 0$ and $|f| < f_{cr}$ (curves 1-2), the pronounced minima at $T = T_{IC}(f)$ appears at $|f| \geq f_{cr}$ (curves 3-4). The sound velocity can be measured in ferroelectrics by Brillouin-scattering and ultrasonic pulse-echo method [8, 18, 21, 24]. In particular the temperature dependencies of sound velocity and attenuation was measured by these methods in the proper uniaxial Sn$_2$P$_2$(Se$_x$S$_{1-x}$)$_6$ ferroelectrics in the vicinity of Lifshitz point (x=0.28 and $T_{LP}$=284 K) [8, 24]. The pronounced softening of A-mode, as remarkable lowering of hypersound velocity for the transverse acoustic phonons that are polarized in crystallographic plane containing the vector of spontaneous polarization and the wave vector of modulation, was observed [see **Fig.4(b)**]. It was shown that elastic softening in the hypersound range is induced mostly by linear interaction between soft optic and acoustic phonon branches [8]. Note that Landau-Khalatnikov model accounting for the modes interaction explains [8] quantitatively temperature dependence of hypersound velocity in the ferroelectric phase.

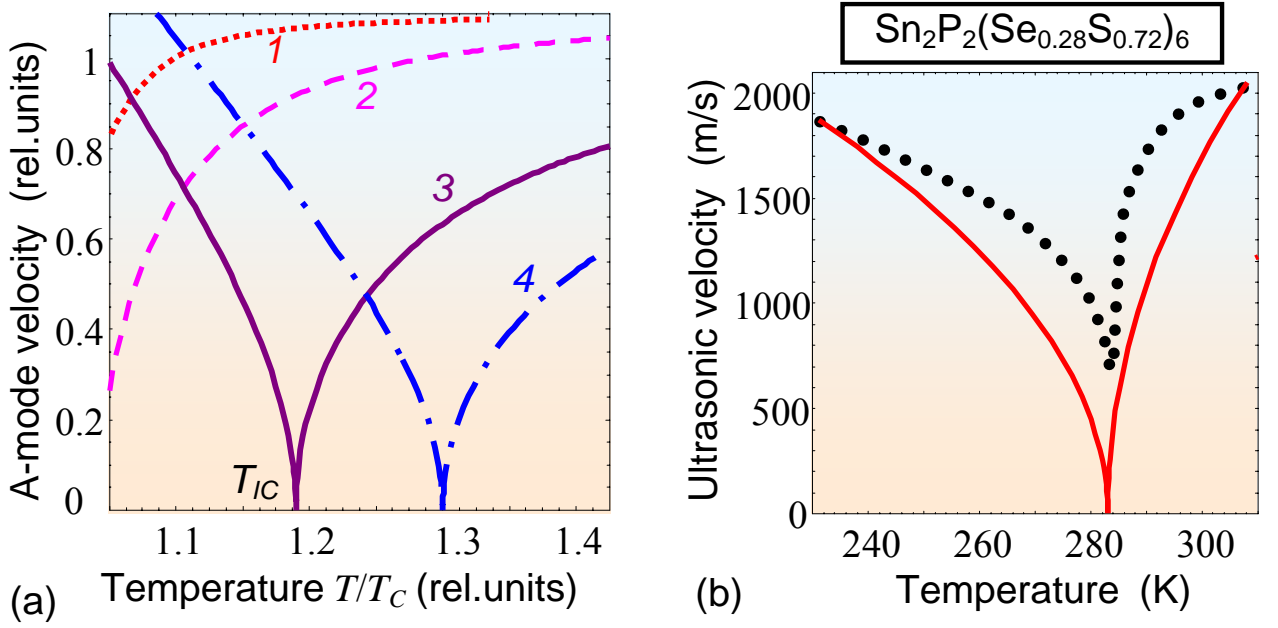

**FIG. 4. (a)** Temperature dependences of the acoustic mode velocity $V(T)$ calculated at $f = 0$ (red curve 1), $|f| = 0.5 f_{cr}$ (magenta curve 2), $|f| = f_{cr}$ (dark violet curve 3) and $|f| \approx 1.25 f_{cr}$ (blue curves 4). **(b)** Experimental temperature dependence of shear XZ ultrasonic mode velocity (x-displacement in z-direction of propagation) near the LP in $Sn_2P_2(Se_{0.28}S_{0.72})_6$ crystal (circles from Ref.[8]) and calculated $V(T)$ at $|f| = f_{cr}$ (solid curve).

## IV. TEMPERATURE DEPENDENCE OF THE DIELECTRIC SUSCEPTIBILITY

Let us proceed for consideration of dielectric susceptibility temperature dependence in SMP. From Eq.(5) the temperature-dependent static dielectric susceptibility, $\tilde{\chi}(k, \omega = 0, T)$, can be represented in the form:

$$\tilde{\chi}(k,0,T) = \frac{vk^2 + c}{\left(\alpha_T(T - T_C) + 3\beta^* P_S^2(T) + 5\gamma P_S^4(T)\right)c + \left(cg + \alpha_S(T)v - f^2\right)k^2 + gvk^4}. \tag{17a}$$

Depending on the temperature and wave vector Eq.(17a) can be represented in the form:

$$\tilde{\chi}(k,0,T) = \begin{cases} \dfrac{vk^2 + c}{gv\left(k^2 - \left(k_0^{cr}\right)^2\right)^2}, & T = T_{IC}, \quad k_{1,2}^{cr} = k_0^{cr} \equiv \sqrt[4]{\dfrac{c\alpha_T(T_{IC} - T_C)}{4vg}}, \\ \dfrac{vk^2 + c}{gv\left(k^2 - \left(k_1^{cr}(T)\right)^2\right)\left(k^2 - \left(k_2^{cr}(T)\right)^2\right)}, & T \neq T_{IC}, \quad k_1^{cr}(T) \neq k_2^{cr}(T). \end{cases} \tag{17b}$$

The x-dependent static susceptibility $\chi(x, T)$ can be obtained after inverse Fourier transformations, which results depends on the temperature via the properties of $k_1^{cr}(T)$ and $k_2^{cr}(T)$:

$$\chi(x,T) = \begin{cases} \left(\dfrac{c}{gv} + \dfrac{1}{g}\dfrac{\partial^2}{\partial x^2}\right)\left(\dfrac{\exp(-ik_1^{cr}|x|)}{ik_1^{cr}(T)} - \dfrac{\exp(-ik_2^{cr}|x|)}{ik_2^{cr}(T)}\right)\dfrac{-1}{(k_2^{cr})^2 - (k_1^{cr})^2}, & T > T_{IC} \text{ or } T < T_C, \\ & \operatorname{Im}(k_1^{cr}) < 0, \operatorname{Im}(k_2^{cr}) < 0 \\[2ex] \left(\dfrac{c}{gv} + \dfrac{1}{g}\dfrac{\partial^2}{\partial x^2}\right)(\sin(k_0^{cr} x) - k_0^{cr} x \cos(k_0^{cr} x))\dfrac{\operatorname{sign}(x)}{2(k_0^{cr})^3}, & T = T_{IC}, \quad k_0^{cr} = \sqrt[4]{\dfrac{c\alpha_T(T_{IC} - T_C)}{4vg}}, \\[2ex] \left(\dfrac{c}{gv} + \dfrac{1}{g}\dfrac{\partial^2}{\partial x^2}\right)\left(\dfrac{\sin(k_1^{cr} x)}{k_1^{cr}(T)} - \dfrac{\sin(k_2^{cr} x)}{k_2^{cr}(T)}\right)\dfrac{\operatorname{sign}(x)}{(k_2^{cr})^2 - (k_1^{cr})^2}, & T_C < T < T_{IC}, \quad k_1^{cr} \ne k_2^{cr}, \\ & \operatorname{Im}(k_1^{cr}) = \operatorname{Im}(k_2^{cr}) = 0 \end{cases} \quad (18)$$

The average susceptibility of a bulk ferroelectric, $\langle\chi(x,0,T)\rangle = \int_{-\infty}^{\infty} dx\,\chi(x,0,T)$, can be derived from Eq.(12) in the temperature ranges of paraelectric ($T > T_{IC}$) and homogeneous ferroelectric phases ($T < T_C$), namely:

$$\langle\chi(T)\rangle = \frac{c}{gv}\frac{1}{(k_2^{cr})^2(k_1^{cr})^2} \equiv \begin{cases} \dfrac{1}{\alpha_T(T - T_C)}, & T > T_{IC}, \\[2ex] \dfrac{1}{\alpha_T(T - T_C) + 3\beta^* P_S^2(T) + 5\gamma P_S^4(T)}, & T < T_C. \end{cases} \quad (19)$$

The derivation of Eq.(19) uses the identities, $\left(\dfrac{1}{(k_1^{cr})^2} - \dfrac{1}{(k_2^{cr})^2}\right)\dfrac{1}{(k_2^{cr})^2 - (k_1^{cr})^2} = \dfrac{1}{(k_2^{cr})^2(k_1^{cr})^2}$ and $(k_2^{cr})^2(k_1^{cr})^2 = \dfrac{c}{vg}\left(\alpha_S - 4\dfrac{q^2}{c}P_S^2\right)$, the latter of which follows from the definition of $k_{1,2}^{cr}$, Eq.(10).

In the SMP phase ($T_C < T < T_{IC}$) the integral $\int_{-\infty}^{\infty} dx\,\chi(x,0,T)$ does not exist is the sense of normal convergence. The integration over the size of the ferroelectric $\int_{-2L}^{2L} dx\,\chi(x,0,T)$ can be performed and results:

$$\langle\chi(L,T)\rangle = \begin{cases} \dfrac{2c}{gv}\dfrac{(k_2^{cr})^2\sin^2(k_2^{cr}L) - (k_1^{cr})^2\sin^2(k_1^{cr}L)}{((k_2^{cr})^2 - (k_1^{cr})^2)(k_2^{cr})^2(k_1^{cr})^2}, & T_C < T < T_{IC}, \\[2ex] \dfrac{2c}{gv(k_0^{cr})^4}(4\sin^2(k_0^{cr}L) - (k_0^{cr}L)\sin(2k_0^{cr}L)), & T = T_{IC}. \end{cases} \quad (20a)$$

Using the equalities $(k_2^{cr})^2(k_1^{cr})^2 \equiv \dfrac{c}{vg}(\alpha_T(T - T_C) + 3\beta^*\delta P^2(T) + 5\gamma\delta P^2(T))$ and $(k_0^{cr})^4 \equiv \dfrac{c\alpha_T(T_{IC} - T_C)}{4vg}$ valid at zero constant component of polarization $P_S^2 = 0$ and its nonzero modulation with the amplitude $\delta P^2(T)$, and performing some averaging over $L$ of the oscillated functions, $\langle\sin^2(k_2^{cr}L)\rangle = \langle\sin^2(k_2^{cr}L)\rangle = 1/2$ and $\langle\sin(2k_0^{cr}L)\rangle = 0$, one get from Eqs.(20a) that

$$\langle\langle\chi(T)\rangle\rangle = \begin{cases} \dfrac{1}{\alpha_T(T-T_C)+3\beta^*\delta P^2(T)+5\gamma\delta P^2(T)}, & T_C < T < T_{IC}, \\ \dfrac{1}{\alpha_T(T_{IC}-T_C)}, & T = T_{IC}. \end{cases} \quad (20b)$$

Hence one can get from Eqs.(19) and (20b) the susceptibility in the form

$$\langle\langle\chi(T)\rangle\rangle \cong \begin{cases} \dfrac{1}{\alpha_T(T-T_C)+3\beta^* P_S^2(T)+5\gamma P_S^4(T)}, & T \leq T_C, \\ \dfrac{1}{\alpha_T(T-T_C)+3\beta^*\delta P^2(T)+5\gamma\delta P^2(T)}, & T_C < T \leq T_{IC}, \\ \dfrac{1}{\alpha_T(T-T_C)}, & T > T_{IC}. \end{cases} \quad (21)$$

Where $P_S^2 = \dfrac{1}{2\gamma}\left(\sqrt{\beta^{*2}-4\gamma\alpha_T(T-T_C)}-\beta^*\right)$ is the spontaneous polarization in FE phase, and $\delta P^2(T) \cong \delta P_0^2(1-T/T_{IC})(T/T_C-1)$ is the polarization modulation amplitude existing in SMP phase only, at that the transition temperature to the SMP phase, $T_{IC}(f) = T_C\left(1+\dfrac{cg}{\alpha_T T_C v}\left(\dfrac{|f|}{\sqrt{cg}}-1\right)^2\right)$, depends on the flexo-constant $f$ in accordance with Eq.(10). The polarization modulation amplitude $\delta P_0^2$ should be found from the free energy minimization in a self-consistent manner, it is also temperature-dependent, but not in a critical way.

**Figure 5(a-b)** illustrates the temperature dependences of direct and inverse average susceptibility calculated from Eq.(21) for the ferroelectric with the second order phase transition from PE to FE phase ($\beta^* > 0$, $\gamma \geq 0$) at different flexocoupling constants $f^*$. Dashed black curves corresponding to $f^* = 0$ are the reference curves, which in fact are conventional Curie-Weiss dependences. The dependences $\langle\langle\chi(T)\rangle\rangle$ and $\langle\langle\chi(T)\rangle\rangle^{-1}$ does not change from the ones calculated at $f^* = 0$ until $|f^*| < f_{cr}^*$. Solid curves corresponding to $|f^*| > f_{cr}^*$ show the small fractures (or "breaks") at $T = T_{IC}(f)$ that becomes more pronounced with $f$ increase (compare solid curves 2-4 in **Fig.5(c)**). The SMP transition temperature changes from $1.17T_C$ to $1.29T_C$ with $f^*$ increase from 4 to 5 n accordance with Eq.(10).

The existence of fracture instead of maxima at $T = T_{IC}$ is characteristic feature of SMP transition [43]. Indeed, results shown in **Fig.5(c)** are in a agreement with experimental data in $Sn_2P_2(S,Se)_6$ [55] that gives us additional background to predict flexo-coupling induced soft acoustic "amplitudon-type" mode in the SMP phase.

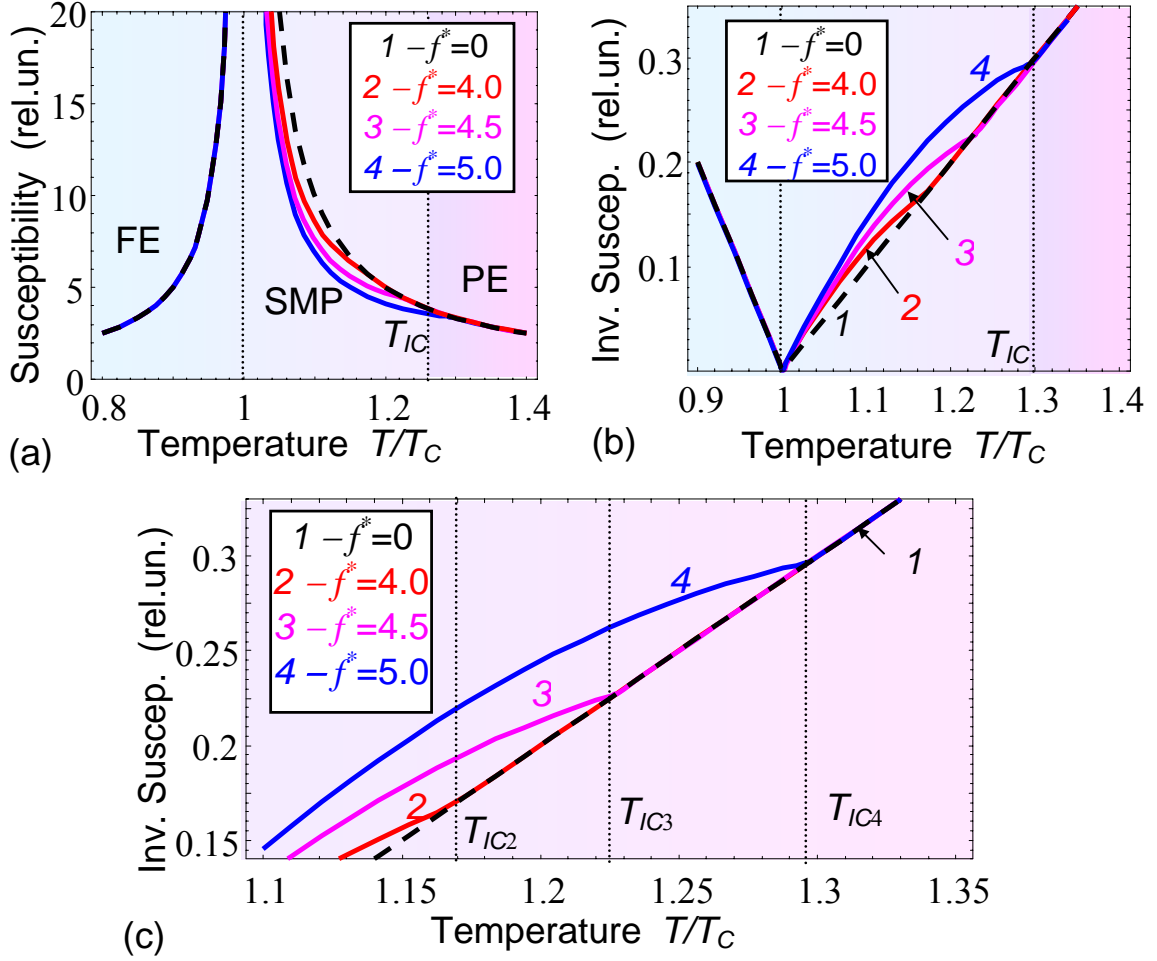

**FIG. 5.** Temperature dependence of the direct **(a)** and inverse **(b,c)** static dielectric susceptibility for a ferroelectric with the second order phase transition calculated for different flexocoupling constants $f^* = 0$ (dashed black curves 1), 4 (red curves 2), 4.5 (magenta curves 3) and 5 (blue curves 4); $v = 10^{-7}$ V s$^2$/m$^2$ and other parameters corresponding to Sn$_2$P$_2$S$_6$ [listed in **Table SI**]. Plot **(c)** is a zoomed region of the plot **(b)** in the vicinity of PE-SMP phase transition.

## VI. CONCLUSION

Using the LGD theory and scalar approximation, we derived analytical expressions for the singular points (zeros, breaks, etc) of the phonon A-mode frequency $\omega(k)$ in dependence on the wave vector $k$ and examined the conditions of the soft acoustic modes appearance in ferroelectrics depending on the magnitude of the flexoelectric coefficient $f$ and temperature $T$ of the ferroelectric.

If the magnitude of the flexocoefficient $f$ is equal to the temperature-dependent critical value $f^{cr}(T)$ at the temperature $T = T_{IC}$, $|f| = f^{cr}(T_{IC})$, then the A-mode frequency tends to zero at $k \to k_0^{cr}$ according to the linear law $\omega(k \to k_0^{cr}) \sim |k - k_0^{cr}|$ and the ferroelectric polarization becomes spatially modulated.

When the magnitude of the flexo-coefficient is more than the critical value $|f| > f^{cr}(T)$ in a temperature range $T_C < T < T_{IC}$ corresponding to the spatially modulated incommensurate phase, the A-mode becomes zero for two wave vectors $k = k_{1,2}^{cr}$ according to the squire root law, $\omega(k \to k_{1,2}^{cr}) \sim \sqrt{|k_{1,2}^{cr} - k|}$, and does not exist in the range of wave vectors $k_1^{cr} < k < k_2^{cr}$. At fixed flexocoefficient $f$ the transition into the spatially modulated phase can appear at the temperature $T_{IC}$ that depends on $f$ as $T_{IC}(f) - T_C \sim (|f| - \sqrt{cg})^2$, where $T_C$ is the ferroelectric Curie temperature.

Since the squire root law has a functional similarity with the temperature dependence of the soft optical phonons near Curie temperature $T_C$, $\omega(k=0,T) \sim \sqrt{T_C - T}$, we stated that the flexocoupling can induce the soft acoustic mode with a $T_{IC}$-dependent frequency $\omega(k \to k_0^{cr}, T) \sim \sqrt{T_{IC} - T_C} |k - k_0^{cr}|$.

The available experimental data on neutron scattering in organic ferroelectric $(CH_3)_3NCH_2COO \cdot CaCl_2 \cdot 2H_2O$ [16] are in semi-quantitative agreement with our theoretical results. For improvement and for quantification of the theory, it is necessary to measure the frequency dependence of the A-mode in a uniaxial ferroelectric with a spatially modulated phase in the temperature interval near its occurrence.

Temperature dependence of the direct and inverse static dielectric susceptibility are in a qualitative agreement with experimental data in $Sn_2P_2Se_6$ [55] that gives us additional background to predict flexocoupling induced soft acoustic "amplitudon-type" mode in the SMP phase. In addition we predicted the appearance of the "rippled" flexocoupling-induced SMP-FE phase in the initially commensurate ferroics. Allowing for the presence of the almost symmetric maxima of dielectric permittivity at $T_{IC}$, the experimental data [55] speaks in favor of the phase existence in $Sn_2P_2(Se_xS_{1-x})_6$ near the boundary of normal FE phase.

# SUPPLEMENT

**Results analyses in dimensionless variables**

In order to quantify the results shown in **Figs.1-3**, let us introduce the dimensionless frequency $\omega^*$, wave vector $k^*$ and parameters $a^*$, $F^*$, $\alpha_S^*$, $Q^*$, $M^*$ and $\mu^*$:

$$\omega^* = \frac{\sqrt{4\nu\rho}}{c}\omega, \qquad k^* = \frac{ak}{\pi}, \qquad a^* = \frac{a}{\pi}\sqrt{\frac{c}{2\nu}}, \qquad F^* = \frac{f^2}{cg}, \qquad \text{(S.1a)}$$

$$\alpha_S^* = \frac{\nu}{cg}\left(\alpha + \left(3\beta - 2\frac{q^2}{c}\right)P_S^2 + 5\gamma P_S^4\right), \quad Q^* = 4\frac{q^2 P_S^2}{c\alpha_S}, \quad M^* = \frac{cM}{2\rho f} \quad \mu^* = \frac{c\mu}{2g\rho}. \quad \text{(S.1b)}$$

where $a$ is the lattice constant. Using the methodology described in details in Ref.[41], we can write the dispersion law (5) for $\omega(k)$ in dimensionless variables (S.1):

$$\omega^{*2}(k^*) = \frac{\left(C^* \pm \sqrt{D^*}\right)}{2\left(\mu^* - 2F^* M^{*2}\right)}, \qquad \text{(S.2a)}$$

$$C^* = 2\alpha_v^* + \left(\frac{k^*}{a^*}\right)^2 (2\mu^* + 1 - 4F^* M^*) + \left(\frac{k^*}{a^*}\right)^4 \mu^*, \qquad \text{(S.2b)}$$

$$D^* = C^{*2} + 4\left(\frac{k^*}{a^*}\right)^2 (2F^* M^{*2} - \mu^*)\left(4\alpha_v^*(1-Q^*) + 2\left(\frac{k^*}{a^*}\right)^2 (1 - F^* + \alpha_v^*) + \left(\frac{k^*}{a^*}\right)^4\right). \quad \text{(S.2c)}$$

In the dimensionless variables the critical value of the flexocoupling constant for which the frequency of acoustic mode becomes zero is $F^* = F_{cr}^*$, where $F_{cr}^*(\alpha_v^*) = 1 + \alpha_v^* + 2\sqrt{\alpha_v^*(1-Q^*)}$. Dimensionless expression for the critical value of wave vector is $k_{cr}^* = a^*\sqrt{(F^* - 1 - \alpha_v^*)}$. At $F^* > F_{cr}^*$ there are two values of the wave vector corresponding to zero A-mode frequency, $k_{1,2}^{*cr} = a^*\sqrt{\left(F^* - 1 - \alpha_v^* \pm \sqrt{(F^* - 1 - \alpha_v^*)^2 - 4\alpha_v^*(1-Q^*)}\right)}$. A mode frequency is nonzero in the regions $0 < k < k_1^{*cr}$ and $k > k_2^{*cr}$. One can expect the appearance of the spatially modulated phase in the "gap" $k_1^{*cr} \le k \le k_2^{*cr}$.

**TABLE SI.** Description of the symbols in the LGD free energy (1a) and kinetic energy (1b), and their numerical values for several ferroelectrics with $g>0$ and $g<0$.

| Description | Symbol and dimension | Incipient and proper ferroelectrics | | | |
|---|---|---|---|---|---|
| | | $Sn_2P_2S_6$ | $Sn_2P_2Se_6$ | | |
| **Coefficient at $P^2$** | $\alpha(T)$ ($\times C^{-2}\cdot mJ$) | $\alpha_T(T - T_C)$ | $\alpha_T(T - T_C)$ | | |
| **Inverse Curie-** | $\alpha_T$ | 16 | 26 | | |

| **Weiss constant** | ($\times 10^5$ C$^{-2}$·mJ/K) | | | | |
|---|---|---|---|---|---|
| **Curie temperature** | $T_C$ (K) | 337 | 193 | | |
| **LGD-coefficient at $P^4$** | $\beta$ ($\times 10^8$ JC$^{-4}$·m$^5$) | +7.42 | −4.8 | | |
| **LGD-coefficient at $P^6$** | $\gamma$ ($\times 10^9$ JC$^{-6}$·m$^9$) | 35 | 85 | | |
| **Electrostriction coefficient** | $q$ ($\times 10^9$ Jm/C$^2$) | 4 (reference interval 1.6 – 4.7) | 5 (defined from acoustic mode tilt) | | |
| **Elastic stiffness coefficient** | $c$ ($\times 10^{10}$ Pa) | 1.6 since $c_{44}$=1.6±0.3 | 1.6 (defined from acoustic mode tilt) | | |
| **Gradient coefficient at $(\nabla P)^2$** | $g$ ($\times 10^{-10}$ C$^{-2}$m$^3$J) | 0.5 (fitting parameter) | − 0.6 (fitting parameter) | | |
| **Gradient coefficient at $(\nabla^2 P)^4$** | $w$ ($\times 10^{-29}$ Jm$^5$/C$^2$) | 1.8 (reference value 1.8) | 2.5 (reference value 2.2) | | |
| **Elastic strain gradient $(\nabla u)^2$** | $v$ ($\times 10^{-9}$ V s$^2$/m$^2$) | 5 (fitting parameter) | 1 (fitting parameter) | | |
| **Static flexo-coefficient** | $f$ (V) | $f_{44}$=±(1.6 - 1.8) (fitting parameter) | ±1.0 (fitting parameter) | | |
| **Dynamic flexo-coefficient** | $M$ ($\times 10^{-8}$ Vs$^2$/m$^2$) | ±2.5 (fitting parameter) | ±1.5 (fitting parameter) | | |
| **Kinetic coefficient** | $\mu$ ($\times 10^{-18}$ s$^2$mJ) | 11.0 (fitting parameter) | 14.5 (fitting parameter) | | |
| **Material density at normal conditions** | $\rho$ ($\times 10^3$ kg/m$^3$) | 1.801 | 2.547 | | |
| **Lattice constant** | $a$ (nm) | $a_x$≈0.93, $a_y$≈0.75, $a_z$≈0.65 at (200 – 350) K | $a_x$≈0.97, $a_y$≈0.77, $a_z$≈0.68 at (200 – 350) K | | |

## References

[1] M. E. Lines and A. M. Glass, *Principles and Application of Ferroelectrics and Related Materials* (Clarendon Press, Oxford, 1977).

[2] W. Cochran, "Crystal stability and the theory of ferroelectricity." Phys. Rev. Lett. **3**(9), 412 (1959)

[3] A. Bussmann-Holder, H. Büttner, and A. R. Bishop. "Coexistence of Polar Order and Local Domain Dynamics in Ferroelectric Perovskites: The Case of SrTi$^{18}$O$_3$", Ferroelectrics, **363:1**, 73 — 78 (2008) http://dx.doi.org/10.1080/00150190802019296

[4] R. A. Cowley. "Lattice Dynamics and Phase Transitions of Strontium Titanate". Phys. Rev. **134**, A981 (1964)

[5] G. Shirane, J. D. Axe, J. Harada, and J. P. Remeika. "Soft ferroelectric modes in lead titanate." Phys. Rev. **B 2**, no. 1: 155 (1970)

[6] A. Bussmann-Holder. "Electron-phonon-interaction-driven anharmonic mode-mode coupling in ferroelectrics: The origin of acoustic-mode anomalies". Phys. Rev. **B, 56**, 10 762 (1997)

[7] A. Bussmann-Holder, H. Buttner, and A. R. Bishop. "Polar-Soft-Mode-Driven Structural Phase Transition in SrTiO$_3$". Phys. Rev. Lett. 99, 167603 (2007)

[8] A. Kohutych, R. Yevych, S. Perechinskii, V. Samulionis, J. Banys, and Yu Vysochanskii. "Sound behavior near the Lifshitz point in proper ferroelectrics". Physical Review **B 82**, no. 5: 054101 (2010).

[9] R. Blinc and B. Zeks, *Soft Mode in Ferroelectrics and Antiferroelectrics* (North-Holland Publishing Company, Amsterdam, Oxford, 1974).

[10] V. G. Vaks, "Introduction to the microscopic theory of ferroelectrics." (Nauka, Moscow, 1973) in russian.

[11] W. Cochran. "Dynamical, scattering and dielectric properties of ferroelectric crystals", Advances in Physics, **18:72**, 157 (1969).

[12] J. D. Axe, J. Harada, and G. Shirane. "Anomalous acoustic dispersion in centrosymmetric crystals with soft optic phonons." Phys. Rev. **B 1**, 3: 1227 (1970).

[13] G. Shirane and Y. Yamada. "Lattice-Dynamical Study of the 110 K Phase Transition in $SrTiO_3$". Physical Review, **177**, (2) 858 (1969).

[14] R. Currat, H. Buhay, C. H. Perry, and A. M. Quittet. "Inelastic neutron scattering study of anharmonic interactions in orthorhombic $KNbO_3$." Phys. Rev. **B 40**, 16: 10741 (1989).

[15] I. Etxebarria, M. Quilichini, J. M. Perez-Mato, P. Boutrouille, F. J. Zuniga, and T. Breczewski. "Inelastic neutron scattering investigation of external modes in incommensurate and commensurate $A_2BX_4$ materials." Journal of Physics: Condensed Matter **4**, no. 44 (1992): 8551.

[16] J. Hlinka, M. Quilichini, R. Currat, and J. F. Legrand. "Dynamical properties of the normal phase of betaine calcium chloride dihydrate. I. Experimental results." Journal of Physics: Condensed Matter **8**, no. 43: 8207 (1996).

[17] J. Hlinka, S. Kamba, J. Petzelt, J. Kulda, C. A. Randall, and S. J. Zhang. "Origin of the "Waterfall" effect in phonon dispersion of relaxor perovskites." *Phys. Rev. Lett.* **91**, 10: 107602 (2003).

[18] . V. Goian, S. Kamba, O. Pacherova, J. Drahokoupil, L. Palatinus, M. Dusek, J. Rohlıcek, M. Savinov, F. Laufek, W. Schranz, A. Fuith, M. Kachlık, K. Maca, A. Shkabko, L. Sagarna, A. Weidenkaff, and A. A. Belik. "Antiferrodistortive phase transition in $EuTiO_3$". Phys. Rev. **B 86,** 054112 (2012).

[19] A. K. Tagantsev, K. Vaideeswaran, S. B. Vakhrushev, A. V. Filimonov, R. G. Burkovsky, A. Shaganov, D. Andronikova, A. I. Rudskoy, A. Q. R. Baron, H. Uchiyama, D. Chernyshov, A. Bosak, Z. Ujma, K. Roleder, A. Majchrowski, J.-H. Ko & N. Setter. "The origin of antiferroelectricity in $PbZrO_3$." *Nature communications* **4**, article number **2229** (2013).

[20] Jong-Woo Kim, P. Thompson, S.Brown, P. S. Normile, J. A. Schlueter, A. Shkabko, A. Weidenkaff, and P. J. Ryan. "Emergent Superstructural Dynamic Order due to Competing Antiferroelectric and Antiferrodistortive Instabilities in Bulk $EuTiO_3$". Phys. Rev. Lett. **110**, 027201 (2013)

[21] R. G. Burkovsky, A. K. Tagantsev, K. Vaideeswaran, N. Setter, S. B. Vakhrushev, A. V. Filimonov, A. Shaganov et al. "Lattice dynamics and antiferroelectricity in PbZrO 3 tested by x-ray and Brillouin light scattering." Phys. Rev. **B 90**, 144301 (2014).

[22] J. Hlinka, I. Gregora, and V. Vorlıcek."Complete spectrum of long-wavelength phonon modes in Sn2P2S6 by Raman scattering". Phys.Rev. **B 65**, 064308 (2002)

[23] Anton Kohutych, Ruslan Yevych, Sergij Perechinskii, and Yulian Vysochanskii. "Acoustic attenuation in ferroelectric $Sn_2P_2S_6$ crystals." Open Physics **8**, 6: 905-914 (2010). DOI: 10.2478/s11534-010-0016-x

[24] Yu M. Vysochanskii, , A. A. Kohutych, A. V. Kityk, A. V. Zadorozhna, M. M. Khoma, and A. A. Grabar. "Tricritical Behavior of $Sn_2P_2S_6$ Ferroelectrics at Hydrostatic Pressure". Ferroelectrics 399, no. 1: 83-88 (2010).http://dx.doi.org/10.1080/00150193.2010.489866

[25] R. M. Yevych, Yu. M. Vysochanskii, M. M. Khoma and S. I. Perechinskii. "Lattice instability at phase transitions near the Lifshitz point in proper monoclinic ferroelectrics". J. Phys.: Condens. Matter **18**, 4047–4064 (2006) (doi:10.1088/0953-8984/18/16/011)

[26] V. S. Mashkevich and K. B. Tolpygo, "The interaction of vibrations of nonpolar crystals with electric fields." Zh. Eksp. Teor. Fiz. **31**, 520 (1957).

[27] Sh. M. Kogan. "Piezoelectric effect under an inhomogeneous strain and an acoustic scattering of carriers of current in crystals". Solid State Physics, **5**, 10, 2829 (1963)

[28] A.K. Tagantsev. "Piezoelectricity and flexoelectricity in crystalline dielectrics". Phys. Rev B, **34**, 5883 (1986)

[29] M.D. Glinchuk, A.V. Ragulya, V.A. Stephanovich. Nanoferroics. Dordrecht: Springer; 2013 May 13., p.378

[30] Sergei V. Kalinin and Anna N. Morozovska. "Multiferroics: Focusing the light on flexoelectricity (Comment)". Nature Nanotechnology, 10, 916 (2015) doi:10.1038/nnano.2015.213

[31] "Flexoelectricity in Solids: From Theory to Applications". Ed. by A.K. Tagantsev and P.V. Yudin, World Scientific (2016)

[32] R. Maranganti and P. Sharma. "Atomistic determination of flexoelectric properties of crystalline dielectrics" Phys. Rev. B 80, 054109 (2009).

[33] N. D. Sharma, C. M. Landis, and P. Sharma. "Piezoelectric thin-film superlattices without using piezoelectric materials". J. Appl. Phys. 108, 024304 (2010).

[34] A. Y. Borisevich, E. A. Eliseev, A. N. Morozovska, C-J. Cheng, J-Y. Lin, Ying-Hao Chu, Daisuke Kan, Ichiro Takeuchi, V. Nagarajan, and S. V. Kalinin. "Atomic-scale evolution of modulated phases at the ferroelectric–antiferroelectric morphotropic phase boundary controlled by flexoelectric interaction". Nature communications **3**, 775 (2012).

[35] E. A. Eliseev, S. V. Kalinin, Yijia Gu, M. D. Glinchuk, V. V. Khist, A. Y. Borisevich, Venkatraman Gopalan, Long-Qing Chen, A. N. Morozovska. "Universal emergence of spatially-modulated structures induced by flexo-antiferrodistortive coupling in multiferroics". Phys.Rev. **B 88**, 224105 (2013)

[36] P. Henning, and E. KH Salje. "Flexoelectricity, incommensurate phases and the Lifshitz point". J. Phys.: Condens. Matter **28**, 075902 (2016).

[37] P V Yudin and A K Tagantsev. "Fundamentals of flexoelectricity in solids". Nanotechnology, **24**, 432001 (2013);

[38] A. Kvasov, and A. K. Tagantsev. "Dynamic flexoelectric effect in perovskites from first-principles calculations." Phys.l Rev. **B 92**, 054104 (2015).

[39] P. Zubko, G. Catalan, A.K. Tagantsev. "Flexoelectric Effect in Solids". Annual Review of Materials Research 43: 387-421. (2013).

[40] Anna N. Morozovska, Yulian M. Vysochanskii, Oleksandr V. Varenyk, Maxim V. Silibin, Sergei V. Kalinin, and Eugene A. Eliseev. "Flexocoupling impact on the generalized susceptibility and soft phonon modes in the ordered phase of ferroics". Phys. Rev. **B 92**, 094308 (2015).

[41] Anna N. Morozovska, Eugene A. Eliseev, Christian M. Scherbakov, and Yulian M. Vysochanskii. "The influence of elastic strain gradient on the upper limit of flexocoupling strength, spatially-modulated phases and soft phonon dispersion in ferroics". Phys. Rev. **B 94**, 174112 (2016).

[42] Eugene A. Eliseev, Anna N. Morozovska, Maya D. Glinchuk, and Sergei V. Kalinin. Missed surface waves in non-piezoelectric solids (http://arxiv.org/abs/1612.08906)

[43] Boris A. Strukov, and Arkadi P. Levanyuk. *Ferroelectric phenomena in crystals: physical foundations*. Springer Science & Business Media, 2012.

[44] C. Kappler, and M. B. Walker. "Symmetry-based model for the modulated phases of betaine calcium chloride dihydrate." Phys. Rev. **B 48**, no. 9 (1993): 5902.

[45] J. Hlinka, M. Quilichini, R. Currat, and J. F. Legrand. "Dynamical properties of the normal phase of betaine calcium chloride dihydrate. II. A semimicroscopic model." Journal of Physics: Condensed Matter **8**, no. 43: 8221 (1996).

[46] S. W. H., Eijt, R. Currat, J. E. Lorenzo, P. Saint-Gregoire, B. Hennion, and Yu M. Vysochanskii. "Soft modes and phonon interactions in $Sn_2P_2S_6$ studied by neutron scattering". The European Physical Journal B-Condensed Matter and Complex Systems **5**, 2: 169-178 (1998).

[47] S. W. H., Eijt, R. Currat, J. E. Lorenzo, P. Saint-Gregoire, S. Katano, T. Janssen, B. Hennion, and Yu M. Vysochanskii. "Soft modes and phonon interactions in Sn2P2Se6 studied by means of neutron scattering". Journal of Physics: Condensed Matter **10**, no. 22: 4811 (1998).

[48]. L.D. Landau and E.M. Lifshitz, *Theory of Elasticity. Theoretical Physics, Vol. 7,* Butterworth-Heinemann, Oxford, U.K. (1998).

[49] G.A. Smolenskii, V.A. Bokov, V.A. Isupov, N.N Krainik, R.E. Pasynkov, A.I. Sokolov, *Ferroelectrics and Related Materials* (Gordon and Breach, New York, 1984).

[50] Yu. M. Vysochanskii and V. Yu. Slivka. "Lifshitz point on the state diagram of ferroelectric". Usp. Fiz. Nauk **162**, 139-160 (1992)

[51] Online Supplementary Materials to "A. N. Morozovska, Yu. M. Vysochanskii, O. V. Varenik, M. V. Silibin, S. V. Kalinin, E. A. Eliseev. Flexocoupling impact on the generalized susceptibility and soft phonon modes in the ordered phase of ferroics. Physical Review **B 92** (9), 094308 (2015)".

[52] V.Yu. Korda, S.V.Berezovsky, A.S.Molev, L.P.Korda, V.F.Klepikov. "A possible generalization of the phenomenological theory of phase transitions in type II ferroelectrics with incommensurate phase". Physica **B 407**, 3388–3393 (2012)

[53] V.Yu.Korda, S.V.Berezovsky a, A.S.Molev, L.P.Korda, V.F.Klepikov. "On importance of higher non-linear interactions in the theory of type II incommensurate systems". Physica **B 42**, 531–33(2013)

[54] A. P. Levanyuk, S. A. Minyukov, and A. Cano, "Universal mechanism of discontinuity of commensurate-incommensurate transitions in three-dimensional solids: Strain dependence of soliton self-energy." Phys. Rev. B **66**, 014111 (2002).

[55] K. Z. Rushchanskii, A. Molnar, R. Bilanych, R. Yevych, A. Kohutych, and Yu. M. Vysochanskii, V. Samulionis and J. Banys. "Observation of nonequilibrium behavior near the Lifshitz point in ferroelectrics with incommensurate phase". Phys.Rev. **B 93**, 014101 (2016)

[56] Elementary calculations lead to the expression $\lambda = \dfrac{\left(\alpha_s^2 \rho v + 4q^2 P_S^2 (c\mu - 2fM + g\rho) - \alpha_S \left(c^2\mu - 2fcM + f^2\rho\right)\right)}{\alpha_S \rho \left(\alpha_S c - 4q^2 P_S^2\right)}$